\documentclass[%
reprint,
 amsmath,amssymb,
aip,jcp,
]{revtex4-2}

\usepackage{graphicx}
\usepackage{dcolumn}
\usepackage{bm}
\usepackage[mathlines]{lineno}
\usepackage{tikz}

\usepackage{mathtools} 

\definecolor{marked}{rgb}{0,0,0} 
\definecolor{hao}{rgb}{0,0,0}
\definecolor{erb}{rgb}{0,0,0}

\begin{document}


\title{Stochastic exciton-scattering theory of optical lineshapes:
Renormalized many-body contributions}

\author{Hao Li}
\email{hli36@central.uh.edu}
\affiliation{Department of Chemistry, University of Houston, Houston, Texas 77204, United~States}

\author{S. A. Shah}
\affiliation{Department of Chemistry, University of Houston, Houston, Texas 77204, United~States}

\author{Eric~R.~Bittner}
\email{ebittner@central.uh.edu}
\affiliation{Department of Chemistry, University of 
Houston, Houston, Texas 77204, United~States}

\author{Andrei Piryatinski }
\email{}
\affiliation{ Theoretical Division, Los Alamos National Laboratory, Los Alamos, NM, 87545 United~States}

\author{Carlos~Silva-Acu\~na}
\affiliation{School of Chemistry and Biochemistry, Georgia Institute of Technology, 901 Atlantic Drive, Atlanta, GA~30332, United~States}
\affiliation{School of Physics, Georgia Institute of Technology, 837 State Street, Atlanta, GA~30332, United~States}
\affiliation{School of Materials Science and Engineering, Georgia Institute of Technology, North Avenue, Atlanta, GA~30332, United~States}


\date{\today}

\begin{abstract}
Spectral line-shapes provide a window into the local environment coupled to 
a quantum transition in the condensed phase. In this paper, we build upon
a stochastic model to account for non-stationary background processes
produced by broad-band pulsed laser stimulation, as distinguished from those for stationary phonon bath.  
In particular, we 
consider the contribution of  pair-fluctuations arising from 
the full bosonic many-body Hamiltonian within a mean-field approximation, 
treating the coupling to the system as a stochastic noise term. 
Using the It{\^o} transformation, we consider two limiting cases for our model
which lead to a connection between the observed spectral fluctuations 
and the spectral density of the environment. 
In the first case, we consider a Brownian environment and show that this produces spectral dynamics that relax to form dressed excitonic states
and recover an Anderson-Kubo-like form for the spectral correlations. 
In the second case,  we assume that the spectrum
is Anderson-Kubo like, and invert to determine the corresponding background. 
Using the Jensen inequality, we obtain an
upper limit for the spectral density for the 
background. 
The results presented here provide the technical tools for
applying the stochastic model to a broad range of 
problems.
\end{abstract}

\maketitle
\section{Introduction}
A spectroscopic 
measurement of a condensed-phase 
system interrogates both the system
and its surrounding local environment. 
In the statistical sense, the 
background density of states 
coupled to the system being probed 
imparts an uncertainty in the energy of 
the transition.  
According to the Anderson-Kubo model (AK), \cite{Anderson:JPSJ1954,Kubo:JPSJ1954} 
this can be incorporated into the spectral
response function by writing that the transition frequency
has an intrinsic time dependence 
\begin{align}
    \omega(t) = \omega_0 + \delta \omega(t)
\end{align}
where $\omega_0$ is the central (mean) transition frequency and $\delta \omega(t)$
is some time-dependent modulation with $\langle\delta\omega(t)\rangle = 0$.
Lacking detailed knowledge of the environment,
it is reasonable to write the frequency auto-correlation 
function in terms of the deviation about the 
mean, $\Delta$ and a single correlation time, $ \tau_c = \gamma^{-1}$, {\em viz.}
\begin{align}
    \langle \delta\omega(t)\delta\omega(0)
    \rangle = \Delta^2 e^{-|t|/\tau_c}.
    \label{eq:2}
\end{align}
The model has two important limits.\cite{hamm_zanni_2d_spectroscopy_2011}
First, if
\textcolor{marked}{$\Delta/\gamma \ll 1$ }, 
the absorption line shape
takes a Lorenzian functional
form with a homogeneous width determined by the dephasing time 
$T_2 = (\Delta^2/\gamma)^{-1}$. 
On the other hand,
if 
$\Delta/ \gamma \gg 1$, 
the absorption spectrum takes a Gaussian form with a line width 
independent of the correlation time.  In this limit, fluctuations
are slow and the system samples a broad distribution of environmental motions. 
Increasing the rate of the fluctuations (i.e. decreasing the 
correlation time) leads to the effect of motional narrowing
whereby the line width becomes increasingly narrow.\cite{Anderson:JPSJ1954,Kubo:JPSJ1954}

We recently developed a stochastic model for this starting from a full many-body 
description of excitons and exciton/exciton interactions and showed how such effects 
are manifest in both the linear and non-linear/coherent spectral dynamics of a 
system.\cite{doi:10.1063/5.0026467,doi:10.1063/5.0026351}
Within our model, the Heisenberg operators
for the optical excitation are driven by stochastic  equations representing 
the transient evolution of a background population of non-optical excitation 
which interact with the optical mode. 
For this, we define the exciton Hamiltonian (with $\hbar = 1$) as
\begin{align}
     H_0(t) =  \hbar\omega_0 a^\dagger_0a_0 +  \frac{V_0}{2} a^\dagger_0 a^\dagger_0a_0a_0 
    +2  V_0 a^\dagger_0 a_0 N(t).
    \label{eq:27}
\end{align}
Where $N(t)$ is derived by assuming the 
optical bright state with operators $[a_0,a_0^\dagger]=1$ are coupled to an 
ensemble of optically dark $q\ne 0$ excitons 
which in turn evolve according to a quantum Langevin 
equation and we assume that the 
dark background can be written in terms of its population
\begin{align}
    N(t) = \left<\sum_q a^\dagger_q(t) a_q(t) \right>,
\end{align}
where $q$ represents the quasi-momentum. 
In deriving this model,  we also assumed that 
an additional term corresponding to pair creation/annihilation could be dropped from 
consideration.
That term takes the form
\begin{align}
    H_{pair} = \sum_{q\ne 0}\gamma_{q} (a_0^\dagger a_0^\dagger
    a_{q}a_{-q}
    + a_{q}^\dagger a_{-q}^\dagger a_0 a_0)
    \label{modelH}
\end{align}
However, such pair-creation/annihilation 
terms terms may give 
important and
interesting contributions to the spectral lineshape, 
especially in systems in which excitons are formed near
the Fermi energy. 
In such systems, the exciton becomes dressed by 
virtual electron/hole fluctuations about the 
Fermi sea producing spectral shifts and broadening 
of the spectral lineshape. 
Such states are best described as exciton/polarons
whose wave function consists of the bare electron/hole
excitation dressed by electron/hole fluctuations. 

Recent advances towards a more microscopic perspective has been presented by  Katsch et al., in which excitonic Heisenberg equations of motion are used to describe linear excitation line broadening in two-dimensional transition-metal dichalchogenides~\cite{Katsch2020}.
Their results 
indicate exciton-exciton scattering from 
a dark background as a dominant 
mechanism in the power-dependent broadening \textcolor{hao}{due to the excitation-induced dephasing (EID)}
and sideband formation. 
Similar theoretical modelling on this class of materials and their van der Waals bilayers have yielded insight into the role of effective mass asymmetry on EID processes~\cite{Erkensten_EID_2020}. These modelling works highlight the need for microscopic approaches to understand nonlinear quantum dynamics of complex 2D semiconductors, but the computational expense could become considerable if other many-body details such as polaronic effects are to be included~\cite{SrimathKandada2020}. As an alternative general approach, an analytical theory of dephasing in the same vein as Anderson-Kubo lineshape theory but that includes
{\em transient} EID and Coulomb screening effects, would be valuable to extract microscopic detail on screened exciton-exciton scattering from time-dependent nonlinear coherent ultrafast spectroscopy, via direct and unambiguous measurement of the homogeneous excitation linewidth~\cite{siemens2010resonance,bristow2011separating}.

It is worth pointing out that our approach is to account for the quadratic spectroscopic effect of a non-stationary background of pumped excitations rather than that arises from the coupling to a stationary bath of phonon modes \cite{SkinnerHsuJPC1986,ReichmanJCP1996}. Such non-stationary excited states can be achieved by external broad-band laser fields in modern spectroscopy. Of our interest is the bright state dressed by non-equilibrium dark excitons rather than well-studied polaronic effects in thermal equilibrium.

In this work, we consider the effect of 
higher-order background fluctuations on 
the spectral lineshape for a given system.
We do so by attempting to connect the
transient line-narrowing and peak shifts
of a spectral transition to an assumed 
stochastic model for the background 
dynamics.  Our results suggest that
the spectral evolution evident in
time-resolved multi-dimensional 
spectroscopic measurements
of semiconducting systems can be use to 
reveal otherwise dark details of background
excitation processes coupled to the
system.

\section{Theoretical Development}
To pursue the effect of the 
pair-fluctuations, we 
start with the basic form of the Hamiltonian
\begin{align}
    H = \hbar\omega (a^\dagger a +1/2)+ \hbar\gamma(t)(a^\dagger a^\dagger + a a)/2
    \label{eq:13}
\end{align}
where $\gamma(t)$ is the coupling
which we take to be an unspecfied 
stochastic process. Formally, 
we can write that
$\gamma(t) \approx \gamma_{pair}N(t) $ 
where $\gamma_{pair}$ is the
coupling constant and $N(t)$ the 
background population at time $t$. 
As described in our recent papers,
this many-body Hamiltonian
follows directly from a full many-body Hamiltonian under the assumption that the
coupling can be described within a long-wavelength
limit (hence, independent of $k$-vector) 
and within a mean-field theory so that the 
number density of the fluctuations 
enters as a single stochastic variable. 
\textcolor{erb}{
The first assumption is justified using the first Born 
approximation scattering theory 
in which the true interaction potential can be
replaced by another finite-ranged 
potential with the same S-wave ($q=0$)
scattering phase-shift.\cite{born1926quantenmechanik}
The second assumption follows from deriving the
Heisenberg-Langevin equations for the background
operators as coupled to ancillary variables, making the Markov approximation, and then treating them in the 
semi-classical limit as ordinary c-numbers.
\cite{doi:10.1063/5.0026467,doi:10.1063/5.0026351}
}

To proceed, we diagonalize $H$ via unitary transform (c.f. Ref.\citenum{wagner1986unitary}, Sec 9.7)
\begin{align}
    \tilde H = e^{-S}He^{+S}
\end{align}
with $S =\xi( a^2 -  (a^\dagger)^2)/2$. 
Transforming the operators, one obtains
\begin{align}
    \tilde a &= e^{-S}a e^{+S} = a \cosh\xi - a^\dagger \sinh\xi\\
    \tilde a^\dagger &= e^{-S}a^\dagger e^{+S} = a^\dagger \cosh\xi - a \sinh\xi,
\end{align}
where $\xi$ is a variational parameter.
\textcolor{erb}{
Note, that this is accomplished by expanding the 
exponents and using the identities
\begin{align}
    [a,S] &=-\xi a^\dagger\\
    [a^\dagger,S] &= -\xi a.
\end{align}}
The transformed operators can be reintroduced 
into the original $H$ to 
produce
\begin{align}
    \tilde H/\hbar &= 
    (a^\dagger a + 1/2)[\omega \cosh( 2\xi) - \gamma(t) \sinh(2\xi)]\nonumber \\
    &+\frac{1}{2}[(a^\dagger)^2 + a^2][\gamma(t)\cosh(2\xi) - \omega\sinh(2\xi)].
\end{align}
Under this transformation,
$\tilde H$ becomes diagonal 
\begin{align}
    \tilde{H}  =  \hbar\tilde{\omega}(t) \left(\tilde{a}^\dagger \tilde{a} + 1/2\right).
\end{align}
with
\begin{align}
    \tanh(2\xi) = \frac{\gamma}{\omega}.
\end{align}
From this, we obtain a renormalized frequency 
\begin{align}
\tilde\omega(t) &= \sqrt{\omega^2 - \gamma(t)^2}.
\end{align}
However, since $\gamma(t)$ is a stochastic process, we need to derive the underlying \textcolor{hao}{stochastic differential equation (SDE)} for the renormalized harmonic frequency, $\tilde\omega(t)$,  in order to compute
 correlation functions. 

In the regime of weak pair-excitation interaction, $\gamma/\omega \ll 1$, the 
eigen-frequency can be approximated as
\begin{align}
    \nonumber 
    \tilde\omega(t) &= \omega \sqrt{1-(\gamma/\omega)^2} \\
    &\approx \omega\left(1-z(t)/2\right),
    \label{eq:w-eigen}
\end{align}
where $z(t)=\gamma(t)^2/\omega^2$. Therefore, $\sqrt{z}$ represents the coupling strength of the pair-excitation relative to the excitation frequency.

After the unitary transformation we have following commutation relations
\begin{align}
    [\tilde{a},\tilde{a}^{\dagger}] &= 1, \\
    [\tilde{H},\tilde{a}] &= -\hbar\tilde{\omega} \tilde{a}, \\
    [\tilde{H},\tilde{a}^{\dagger}] &= \hbar\tilde{\omega} \tilde{a}^{\dagger},
\end{align}
which lead to the time evolution of the operators 
in the interaction picture
\begin{align}
    \tilde{a}_{\rm I}(t)  &= \tilde{a}_0 \exp\left[-i\int_0^t \tilde{\omega}(\tau){\rm d}\tau\right] \nonumber \\
    &\approx \tilde{a}_0 \exp(-i\omega t) \exp\left[\frac{i\omega}{2} \int_0^t z(\tau) {\rm d}\tau \right],
\end{align}
where $\tilde{a}_0=\tilde{a}(0)$ is the initial condition. Because $\tilde{a}(t)$ commutes at different times, the commutation relation of the dipole operator remains unchanged under the unitary transformation.

\begin{widetext}
For the moment, we leave the 
stochastic variable unspecified and find the linear response function
\begin{align}
    S^{(1)}(t) =& \frac{i}{\hbar} \left< [\hat \mu(t),\hat\mu(0)] \rho(-\infty)\right> \nonumber \\
    =& \frac{i}{\hbar}\mu^2\left<\left[\tilde{a}^{\dagger}(t),\tilde{a}_0\right]\rho(-\infty) - {\rm c.c} \right> \nonumber \\
    =& \frac{2\mu^2}{\hbar} {\rm Im}\left<\exp(i\omega t)\exp\left[-\frac{i\omega}{2}\int_0^t z(\tau){\rm d}\tau \right]\right> \\
    =& \frac{2\mu^2}{\hbar} {\rm Im}\left\{\exp(i\omega t) \exp\left[\sum_{n=1}^{\infty}\frac{(-i\omega/2)^n}{n!}\left<\left(\int_0^t z(\tau){\rm d}\tau\right)^n\right>_{\rm c}\right]\right\}
\end{align}
in the form of cumulant expansion, where $\left<x^n\right>_{\rm c}$ denotes the $n$-th cumulant. 
\textcolor{hao}{According to the theorem of Marcinkiewicz,\cite{Marcinkiewicz1939,MarcinkiewiczTheorem:PRA1974}
the cumulant generating function is a polynomial of degree no greater than two to maintain the positive definiteness of the probability distribution function.} Therefore, we truncate the cumulant expansion to the second order and write the spectral line-shape functions $g_1(t)$ and $g_2(t)$ from the first and second cumulants,
\begin{align}
    g_1(t)=& \int_0^t\langle z(\tau)\rangle {\rm d}\tau 
\end{align}
and 
    \begin{align}
        g_2(t)=& \int_0^t \int_0^t \langle z(\tau),z(\tau')\rangle {\rm d}\tau{\rm d}\tau',
    \end{align}
respectively.

So far we have not limited $z(t)$, equivalently speaking $\gamma(t)^2$, to any particular stochastic process. In principle, once the stochastic differential equation of $\gamma(t)$ is specified, one can find its cumulants thence the mean and covariance of $z(t)$ which determine the spectral line shape functions.   
With the $g_1(t)$ and $g_2(t)$ expressions 
in hand, one can go on to write expressions for 
the higher-order spectral response terms as in Ref.~\citenum{doi:10.1063/5.0026467,doi:10.1063/5.0026351}.   
Our general procedure is to first define the SDE for either the background $\gamma(t)$ process or the phenomenological driven $z(t)$ process, use the It{\^o} identity to determine the 
SDE for the frequency as a 
transformed process under $z(t)=\gamma^2(t)/\omega^2$,
either analytically or numerically determine the mean and covariance, and finally 
compute the cumulants needed for the spectral responses.

\subsection{Integrating the stochastic variables}
In order to compute the spectral responses,  we need 
to specify underlying SDE that gives rise to $\gamma(t)$.
In the most general case, 
\begin{align}
    {\rm d} \gamma(t) = A[\gamma] {\rm d}t + B[\gamma] {\rm d}W_t,
\end{align}
where $W_t$ is a Wiener process. We can write the SDE for $\gamma(t)^2$ in using the It{\^o} formula
\begin{align}
    \nonumber
    {\rm d}\left[\gamma(t)^2\right] &= \frac{\partial \gamma^2}{\partial t} {\rm d}t + \frac{\partial \gamma^2}{\partial \gamma} {\rm d}\gamma +\frac{1}{2} \frac{\partial^2\gamma^2}{\partial \gamma^2}\left({\rm d}\gamma\right)^2 \\
    &\approx \left(2\gamma A[\gamma]+B[\gamma]^2\right){\rm d}t + 2\gamma B[\gamma]{\rm d}W_t.
    \label{eqn:SDE_gamma2}
\end{align}
In terms of $z(t)=\gamma(t)^2/\omega^2$, the SDE reads
  \begin{align}
    \nonumber 
    \omega {\rm d}z  &= \left\{\frac{2\gamma}{\tilde\omega}A[\gamma] + \left(\frac{\gamma^2}{\tilde\omega^3}+\frac{1}{\tilde\omega}\right)B[\gamma]^2\right\} {\rm d}t + \frac{2\gamma}{\tilde\omega}B[\gamma]{\rm d}W_t \\ \nonumber
    &\approx \left\{\sqrt{z}\left(2+z\right)A[\gamma] + \frac{2+3z+3z^2}{2\omega}B[\gamma]^2\right\}{\rm d}t + \sqrt{z}\left(2+z\right)B[\gamma]{\rm d}W_t \\
    &\approx \left\{2\sqrt{z}A[\gamma] + \frac{1}{\omega}B[\gamma]^2\right\}{\rm d}t + 2\sqrt{z}B[\gamma]{\rm d}W_t. 
  \end{align}
Here we use $(1-z)^{-1}\approx 1+z$ and $(1-z)^{-3}\approx 1+3z$, for $z(t)\ll 1$. 
Keeping only the $\sqrt{z}$ terms, the equation is exactly the same as Eq.(\ref{eqn:SDE_gamma2}).  Hereafter, we are going to neglect $z$ and higher order terms but keep only $\sqrt{z(t)}$ (equivalently, $\gamma$) because $\gamma(t)/\omega \ll 1$ and determines the magnitude of pair-excitation interactions recalling $\gamma(t)=\gamma_{\rm pair} N(t)$. 

In our previous work, the covariance function of $N(t)$ characterizes the exciton-exciton coupling. The effect of multiple exciton interaction may be included in the model by taking into account the autocorrelation function of $\gamma(t)$ higher orders. So hereafter the pair-excitation coupling strength $\gamma(t)^2$ or the relative amplitude $\gamma(t)^2/\omega^2$ should be of our major interest in the stochastic treatment.
Without loss of generality, we consider $\gamma(t)$ as a Gaussian process whose mean is zero and the covariance at any two times is known. Because $\gamma(t)$ is Gaussian, all its moments of order higher than two can be expressed in terms of those of the first and second order. Therefore, we find the mean value and the covariance function of $z(t)$
\begin{align}
    \langle z(t)\rangle &= \frac{1}{\omega^2} {\rm Var}[\gamma(t)], 
    \label{eqn:zmean}\\
    \langle z(t_1),z(t_2)\rangle &= \frac{2}{\omega^4} \left<\gamma(t_1),\gamma(t_2)\right>^2.
    \label{eqn:zcov}
\end{align}

We now examine \textcolor{hao}{two} special cases that 
can be solved exactly.  
First, we consider the case when the coupling $\gamma(t)$ follows a mean-reverting (Ornstein-Uhlenbeck) process \textcolor{hao}{with the mean reversion rate $\theta$}. 
Under this assumption, the background fluctuations
have a single characteristic variance and correlation
time such that (for a stationary process)
\begin{align}
    \langle \gamma(t), \gamma(t+\Delta t)\rangle 
    =
    \frac{\sigma^2}{\theta}e^{-\theta |\Delta t|}.
\end{align} 
This is of course the simplest model for the 
fluctuations.
We then consider the case where 
the resulting frequency fluctuations themselves
are mean-reverting.  This latter case 
corresponds to the more typical Kubo-Anderson model.

\subsection{Treating the interaction as a Gauss-Markov process\label{Sec:2.1}}
A key feature of our approach is that the coupling $\gamma_t$
obeys a stochastic differential equation representing 
the density of states of the background.  As a first
approximation, we shall assume 
that $\gamma(t)$ follows from a stationary Gauss-Markov
(Ornstein-Uhlenbeck (OU)) process specified by the stochastic
differential equation
\begin{align}
    {\rm d}\gamma_t = - \theta \gamma_t {\rm d}t + \sigma {\rm d}W_t.
    \label{eq:gamma-OU-SDE}
\end{align}
This case would correspond to the vacuum 
fluctuations about bare exciton state.  We should emphasize that this is not properly in the regime 
of quantum fluctuations since we have not enforced
the bosonic commutation relation within the background. 
Applying the It{\^o} identity, we arrive at a SDE for the 
exciton frequency,
\begin{align}
    {\rm d}z_t  &= 2\theta \left(\frac{\sigma^2}{2\theta\omega^2} - z_t\right){\rm d}t + \frac{2\sigma}{\omega} \sqrt{z_t}{\rm d}W_t,
\end{align}
in which the relaxation rate is $2\theta$, and the drift term $\sigma^2/2\theta\omega^2$ corresponds to the mean value of the stationary state. The formal solution, analogous to $\gamma(t)$ as the solution of the Ornstein-Uhlenbeck SDE, is
\begin{align}
    z(t)^{1/2} &=\left[z(0)\right]^{1/2} e^{-\theta t} + \frac{\sigma}{\omega} \int_0^t e^{-\theta(t-s)} {\rm d}W_s \\
    \nonumber
    \gamma(t) &= \gamma(0) e^{-\theta t} + \sigma \int_0^t e^{-\theta(t-s)} {\rm d}W_s.
\end{align}
Using It{\^o} isometry we find the average
\begin{align}
    \langle z(t)\rangle = z_0 e^{-2\theta t} + \frac{\sigma^2 }{2\omega^2\theta}\left(1-e^{-2\theta t}\right),
    \label{eqn:z-mean}
\end{align}
and the covariance function
  \begin{align}
    \langle z(t),z(s)\rangle = \sigma_{z_o}^2 e^{-2\theta(t+s)} + \frac{\sigma^4}{2\theta^2\omega^4}\left[e^{-\theta|t-s|}-e^{-\theta(t+s)}\right]^2 + \frac{2\sigma^2}{\theta\omega^2}z_0 e^{-\theta(t+s)} \left[e^{-\theta|t-s|}-e^{-\theta(t+s)}\right]
    \label{eqn:zt-cov}
  \end{align}
where $z_0=\langle z(0)\rangle$, and $\sigma_{z_o}^2=\langle\left(z(0)-z_0\right)^2\rangle$ is the variance of the initial condition. In the case of deterministic initial condition $z(0)=z_0$, the first term vanishes. In case of stationary state $s,t \rightarrow +\infty$, the covariance function is determined by the time interval $\Delta t$
\begin{align}
    \langle z(t),z(t+\Delta t)\rangle = \frac{\sigma^2}{2\theta^2\omega^4} e^{-2\theta|\Delta t|}.
    \label{eqn:stationarycov}
\end{align}
Because $\gamma(t)$ is a Gaussian and Markovian process with covariance 
\begin{align}
    \langle\gamma(s),\gamma(t)\rangle = \frac{\sigma^2}{2\theta} \left[e^{-\theta|t-s|} - e^{-\theta(t+s)}\right],
\end{align}
we can write the mean value and covariance of $z(t)$ according to Eqs.~(\ref{eqn:zmean}) and (\ref{eqn:zcov})
\begin{align}
    \langle z(t) \rangle_{t\rightarrow \infty} = \frac{\sigma^2}{2\theta\omega^2}, \label{eq:z-mean-stationary}\\
    \langle z(t),z(t+\Delta t)\rangle _{t\rightarrow \infty} &= \frac{\sigma^4}{2\theta^2\omega^4} e^{-2\theta|\Delta t|},
\end{align}
which agree with the direct solutions of Eqs.~(\ref{eqn:z-mean}) and (\ref{eqn:stationarycov}), respectively.

From these, we arrive at the following expressions of lineshape functions related to the first cumulant
\begin{align}
    g_1(t)=& \int_0^t\langle z(\tau)\rangle {\rm d}\tau \nonumber \\
    =& \frac{\sigma^2 t}{2\theta\omega^2} + \frac{1}{2\theta}\left(z_0-\frac{\sigma^2}{2\theta \omega^2}\right)\left(1-e^{-2\theta t}\right),
    \label{eq:g1-new}
\end{align}
and to the second cumulant
    \begin{align}
        g_2(t)=& \int_0^t \int_0^t \langle z(\tau),z(\tau')\rangle {\rm d}\tau{\rm d}\tau' \nonumber \\
        =& \frac{\sigma_{z_o}^2}{4\theta^2}\left(1-e^{-2\theta t}\right)^2 + \frac{\sigma^4}{8\theta^4\omega^4} \left(e^{-4\theta t} + 8\theta t e^{-2\theta t} + 4e^{-2\theta t} + 4\theta t -5\right) + \frac{\sigma^2}{2\theta^3\omega^2} z_0 \left(1 - 4\theta t e^{-2\theta t} - e^{-4\theta t}\right)
        \label{eq:g2-new}
    \end{align}

\color{erb}{
\subsubsection{Effect on 2D spectroscopy}

The inhomogeneous and homogeneous contributions
to the lineshape can be separated using 2D coherent
spectroscopic methods. ~\cite{doi:10.1021/acs.jpcc.2c00658,fuller2015experimental,cho2008coherent,tokmakoff2000two,bristow2011separating}
In most molecular applications of 2D spectroscopy, the
evolving background plays little to no role in the
spectral dynamics.  However, 
evolving background does affect the spectral lineshape by mixing absorptive and dispersive features in the real and imaginary spectral components. Generally speaking, systems
lacking background dynamics exhibit absorptive line-shapes
and dispersive lineshapes are a consequence of many-body correlations~\cite{doi:10.1063/5.0026351}, consistent with the analysis of similar measurements in semiconductor quantum wells~\cite{Li_EID_2006}.
Furthermore it is useful to compare the model presented
here, which pertains to the exciton/exciton exchange
coupling, versus our previous model which did not include 
this term and only considered the direct (Hartree) interaction. 
For this, we compute the third-order response
\begin{align}
    S^{(3)}(\tau_3,\tau_2,\tau_1)
    = \langle \mu(\tau_3)[\mu(\tau_2),[\mu(\tau_1),[\mu(0),\rho(-\infty]]]\rangle.
\end{align}
where $0\le \tau_1 \le \tau_2 \le \tau_3$ correspond to the 
interactions times of a series of laser pulses. 
This can be evaluated using the double-sided 
Feynman diagram technique,~\cite{mukamel1995principles}
and assuming that the light-matter interaction can be treated
within the impulsive/rotating-wave approximation. 
One easily finds the responses for the various Liouville-space paths take the form
\begin{align}
    R_n(\tau_3,\tau_2,\tau_1) = \left(\frac{i}{\hbar}\right)^3\mu^4
    \left\langle \exp\left[i\sum_{j=1}^3(\pm)_j
    \int_0^{\tau_j}\tilde\omega(\tau){\rm d}\tau\right]
    \right\rangle
    \label{Eq:Rn}
\end{align}
where the angular brackets denote averaging over the 
stochastic noise term and the $(\pm)_j$ corresponds to 
whether or not the time-step involves an excitation (+)
or de-excitation (-) of the system.  
The time-ordering of the three optical pulses in the experiment and phase-matching conditions define the specific excitation pathways, \textcolor{hao}{based on which \textit{photon echo} ($k_s = -k_1+k_2+k_3$) and \textit{virtual echo} ($k_s = +k_1-k_2+k_3$) signals can be obtained by heterodyne detection (the fourth pulse) ~\cite{cho2008coherent}. Equivalently, in the experiments using co-linear phase-modulated pulses, \textit{rephasing} [$-(\phi_{43}-\phi_{21})$] and \textit{non-rephasing} [$-(\phi_{43}+\phi_{21})$] signals can be measured.} In the rephasing experiment, the pulse sequence is such that the phase evolution of the polarization after the first pulse and the third pulse are of opposite sign, while in the non-rephasing experiment, they are of the same sign. 
Eq.(\ref{Eq:Rn}) can be evaluated by cumulant expansion and
the full expressions are given in Appendix~\ref{app:D}.
Since the $\tilde\omega(\tau)$ corresponds to a {\em non-stationary} process, both the lineshape functions $g_1$ and $g_2$ 
contribute to the output signal.

\textcolor{hao}{Fig.~\ref{fig:2d-spectrum} presents the 2D rephasing and non-rephasing spectra corresponding to a single quantum state dressed by the pair-excitation terms. Focusing on the effect of interactions of paired excitations, rather than that of the initial condition, we set $\sigma_{\gamma_o}^2=\sigma^2/(2\theta)$ so that the initial fluctuation is the same as that of the Wiener process\cite{doi:10.1063/5.0026467}. The initial distribution of $z(0)$ can be found from Eqs.(\ref{eqn:z0}) and (\ref{eqn:z0-fluc}) in Appendix \ref{sec:gaussian}.}

\textcolor{hao}{The ``dispersive'' lineshape is observed in the real spectra for both rephasing and non-rephasing pulse sequences, which is a clear indication of the EID. The center of the peak deviates from the bare exciton energy $\hbar\omega=2.35~{\rm eV}$ (black dashed lines) due to the coupling between exciton pairs. Both the absorption and emission energies shift to red because $z(t)$ is positive by definition Eq.(\ref{eq:w-eigen}). Although the Hamiltonian is diagonal after the exciton/polaron transformation using matrix $S$, the diagonal peaks are off the diagonal. Noting Eqs.(\ref{eq:w-eigen}) and (\ref{eq:z-mean-stationary}), we find that the emission frequency shift from $\omega$ by $-\sigma^2/(4\theta\omega)$ (red dashed line), as long as the time scale of the experiment is greater than the relaxation time $(2\theta)^{-1}$. Indeed, this energy discrepancy attributed to stationary state of $z(t)$ can be considered as the exciton/polaron dressing energy. Regarding the absorption frequency measured by the first two pulses, because the system may not have sufficient time to relax, we can estimate, from Eq.(\ref{eqn:z-mean}), that the shift ranges between $\omega z_0/2$ and $\sigma^2/(4\theta\omega)$. The median $\omega z_0/4+\sigma^2/(8\theta\omega)$ is shown as red dashed line for absorption.}


\begin{figure}
    \centering
    \includegraphics{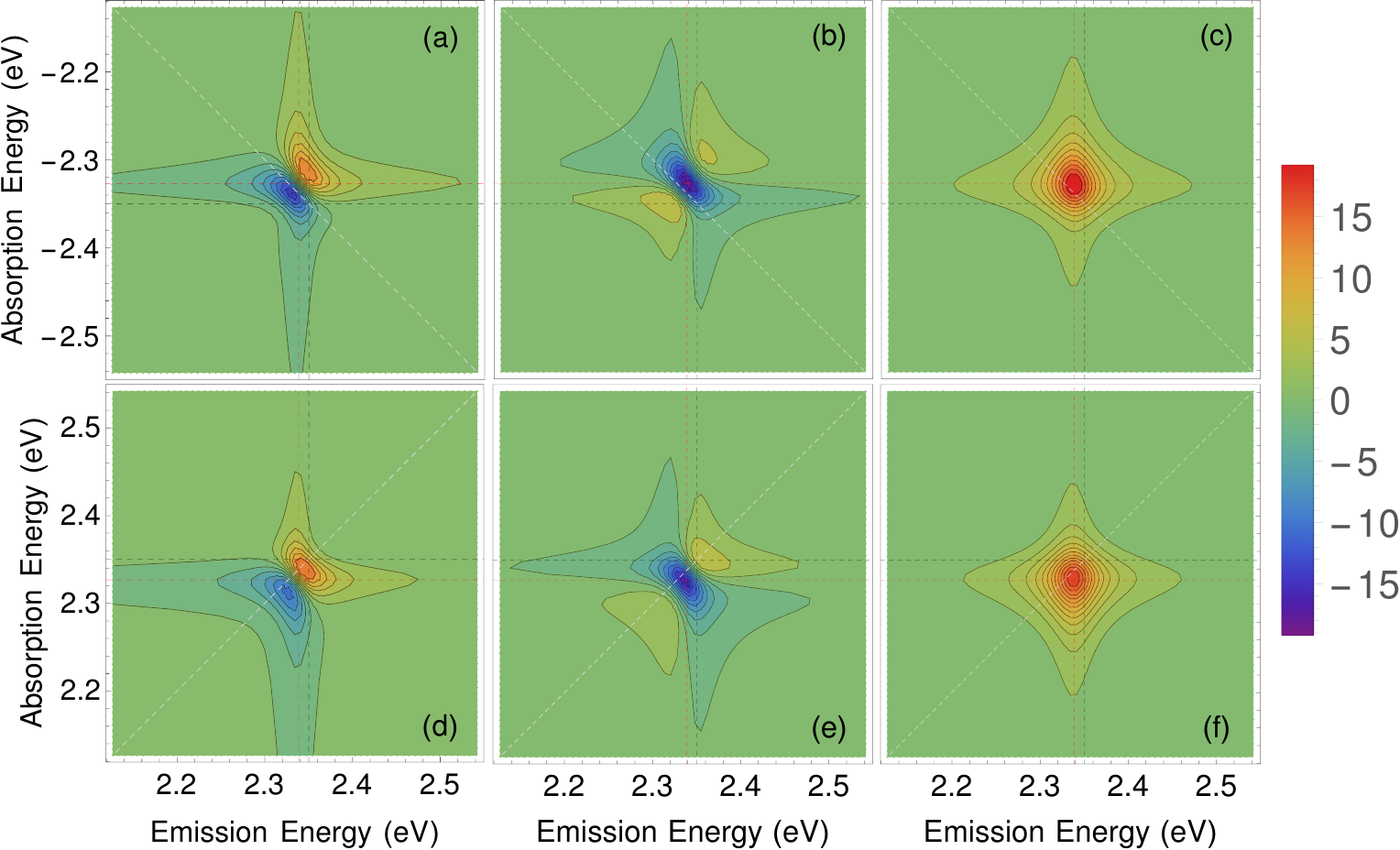}
    \caption{Rephasing (top) and non-rephasing (bottom) spectra at population time $t_2=100$ fs based on the SDE in Eq.(\ref{eq:gamma-OU-SDE}). (a) and (d) are the real, (b) and (e) are the imaginary, (c) and (f) are the norm of the spectra. The parameters used in the simulation: $\hbar\omega=2.35$ eV, $\sigma=0.05~{\rm fs}^{-3/2}$, $\theta=0.01~{\rm fs}^{-1}$, and $\gamma_0=0.5~{\rm fs}^{-1}$.}
    \label{fig:2d-spectrum}
\end{figure}

}

\subsubsection{Comparison to the Anderson-Kubo model and our previous excitation-induced dephasing (EID) theory}
The well-known Anderson-Kubo theory describes the line shape broadening with regard to the stationary state of a random variable (usually the frequency fluctuation, $z_t$ here) characterized by an Ornstein-Uhlenbeck  process. The expansion of the linear optical response function leads to the first cumulant $g_1^{\rm AK}(t)=\mu t$, in which $\mu$ is the drift term, i.e., the mean value, and the second cumulant 
\begin{align}
    g_2^{\rm AK}(t) = \frac{\sigma^2}{2\theta^3} \left( e^{-\theta t} + \theta t - 1\right).
\end{align}

In the short time limit $\theta t \ll 1$, the first cumulant in Eq.~(\ref{eq:g1-new}) turns to $g_1(t) \approx z_0 t$, which has the same linear form as $g_1^{\rm AK}(t)$. It also agrees with the counterpart in our previous publication\cite{doi:10.1063/5.0026467,doi:10.1063/5.0026467} 
\begin{align}
    g_1^{\rm EID}(t)=\frac{z_0}{2\theta}\left(1-e^{-2\theta t}\right)\approx z_0 t.
    \label{eq:g1-EID1}
\end{align}
It is worth noting that the relaxation rate here is $2\theta$ instead of $\theta$, because $\gamma(t)^2$ is the stochastic process of interest rather than $\gamma(t)$ characterized by the rate $\theta$. For deterministic initial condition, $z_0=\gamma_0^2/\omega^2$. The first cumulant $g_1(t)$ results in a red shift of $z_0\omega/2$ in the linear spectrum in the short time limit, which is determined by the initial average of the stochastic process $z(t)$.

When the initial fluctuation of $\gamma(t)$ obeys the same Ornstein-Uhlenbeck  process, we conclude $\sigma_{\gamma_o}^2=\sigma^2/2\theta$ from the stationary state corresponding to the long-time limit where Eq.~(\ref{eqn:z-mean}) turns into ${\rm Var}[\gamma(t)]=\sigma^2/2\theta$. 
Considering Eq.~(\ref{eqn:z0}), we have
\begin{align}
    g_1(t)= \frac{\sigma^2}{2\theta\omega^2}t + \frac{\gamma_0^2}{2\theta\omega^2} \left(1-e^{-2\theta t}\right),
\end{align}
in which the second term looks similar to the $g_1^{\rm EID}(t)$ function in Eq.~(\ref{eq:g1-EID1}). However, $z_0=\gamma_0^2/\omega^2$ is true only for the deterministic initial condition, which is not the case in the above equation. Eq.~(\ref{eq:g1-EID1}) given in our previous publication
leads to a time-dependent red shift that eventually vanishes after sufficiently long time. The first term then can be considered as a correction term that accounts for the interaction of paired-excitation and leads to a constant red shift of $\sigma^2/4\theta\omega$.

Therefore, the first cumulant of the present model produces the red shift similar to but more complex than the counterpart in our previous model, where interactions between paired-excitations are neglected. The initial frequency shift $z_0\omega/2$ agrees with the Anderson-Kubo theory, but converges to $\sigma^2/4\theta\omega$ rather than decaying to zero as in our previous papers.

Regarding the second cumulant, $g_2(t)$, the result from our 
previous work reads
\begin{align}
    g_2^{\rm EID}(t) = \frac{\sigma_{\gamma_o}^2}{\theta^2}\left(1-e^{-\theta t}\right)^2 + \frac{\sigma^2}{2\theta^3}\left(2\theta t + 4e^{-\theta t} -e^{-2\theta t} -3\right).
\end{align}
Compared to Eq.~(\ref{eq:g2-new}), the first term is recovered; 
however, the present model provides a more sophisticated description of the dependency on the initial average $z_0$.

In the limiting case of stationary state where $\sigma_{\gamma_o}^2=\sigma^2/2\theta$, the second cumulant in the present model turns into
\begin{align}
    g_2(t) = \frac{\sigma^4}{4\theta^4\omega^4}\left(e^{-2\theta t} + 2\theta t -1\right) + \frac{\sigma^2 \gamma_0^2}{\theta^3\omega^4}(e^{2\theta t} - 2\theta t -1) e^{-2\theta t},
\end{align}
in which the first term reproduces the Anderson-Kubo lineshape but with half correlation time $\tau_c=(2\theta)^{-1}$ compared to that of the Anderson-Kubo theory $\theta^{-1}$. Furthermore, the second term gives the line broadening due to the initial average of the background exciton population, $\gamma_0^2$, which only results in a frequency shift in our previous model. 


\subsection{Inverting the spectral lineshape to extract the 
background process\label{Sec:2.2}}
\textcolor{erb}{
The practical utility of any spectroscopic method
is to extract information about the system or sample 
being interrogated.  Any inversion approach will
depend upon the model used for the input spectra
and the model used to describe the coupling between 
the system and its environment. 
Here we consider the case in 
which the line-shape function follows from 
the Anderson-Kubo model, but the underlying
background process is due to the pair fluctuation terms.
} 
From a spectroscopic 
point of view, we will have the typical 
motional narrowing and inhomogeneous broadening limits; however, their physical 
origins depend upon actual coupling to 
the background fluctuations. 
For this we consider just the stationary
limit with the goal of relating the
spectral lineshape to the underlying 
spectral density of the pair fluctuations.

Since $\gamma(t) = \omega\sqrt{z(t)}$
and we assume that
$z(t)$ follows from an Ornstein-Uhlenbeck process, one has the SDE of $\gamma(t)$
\begin{align}
    {\rm d}\gamma_t = - \left(\frac{\theta}{2}\gamma_t + \frac{\sigma^2 \omega^4}{8} \gamma_t^{-3}\right) {\rm d}t + \frac{\sigma\omega^2}{2}\gamma_t^{-1} {\rm d}W_t,
\end{align}
with solution
\begin{align}
    \gamma(t)^2=\gamma(0)^2 e^{-\theta t} + \sigma \omega^2 \int_0^t e^{-\theta(t-\tau)} {\rm d}\tau.
\end{align}
Taking this to be a stationary process, we can 
integrate the SDE to find
\begin{align}
    \gamma(t)=\pm \omega \sigma^{1/2} \left[\int_0^t e^{-\theta(t-\tau)} {\rm d}W_{\tau}\right]^{1/2},
\end{align}
and use this to construct the spectral density 
of the underlying many-body dynamics.
\begin{align}
    S(\Omega) =& \int_{-\infty}^{+\infty} {\rm d}t  \langle \gamma_t\gamma_0\rangle e^{-i\Omega t}
\end{align}
However, since this involves taking 
averages over the Wiener process, we can not 
directly use the It{\^o} identity ${\rm d}W^2_t = {\rm d}t$ to
perform the integration.  We can, however, 
find the upper limit of the covariance according to Jensen's inequality which relates the value of a convex  function of an integral to the integral of the convex function\cite{Jensen1906,Inequalities}. 
Here, taking $X$ as a
random variable and $\varphi$ as a convex function,
Jensen's inequality gives
\begin{align}
    \phi(E[X]) \leq E[\phi(X)].
\end{align}
This is essentially a statement that the secant line
of a convex function lies above the graph of the function itself.  
As a corollary, the 
inequality is reversed for a concave function such as $\sqrt{x}$. In cases of stationary state or $\gamma(0)=0$, we have
\begin{align}
    \langle\gamma(t)\rangle \leq \omega\sigma^{1/2}\left<\int_0^t e^{-\theta(t-\tau)}{\rm d}W_{\tau}\right>^{1/2} =0,
\end{align}
which indicates that $\langle\gamma(t)\rangle=0$. The difference between the
left and right sides of the inequality is termed the Jensen gap. 
Employing the inequality over a 
small integration range $\Delta t$
\begin{align}
    \langle \gamma(t+\Delta t)\gamma(t)\rangle \leq \frac{\sigma\omega^2}{\sqrt{2\theta}} e^{-\theta|\Delta t|/2}.
    \label{eq:stationary-cov}
\end{align}
This then implies a spectral density 
of 
\begin{align}
    S(\Omega) =& \int_{-\infty}^{+\infty} {\rm d}t  \langle \gamma_t\gamma_0\rangle e^{-i\Omega t} \nonumber \\
    \approx & \frac{2\sqrt{2\theta}\sigma\omega^2}{\theta^2+\Omega^2}.
    \label{eq:44}
\end{align}
that can be well approximated by a Lorentzian 
in the limit that the Jensen inequality becomes an
equality.  
Since the Lorentzian spectral density 
implies an underlying OU process for $\gamma_t$, in this limit the two 
 cases considered here become identical. 
 The equality is only satisfied
 when the convex (or concave) 
 function is 
 nearly linear over the entire 
 given range of 
 integration which implies that the
 $\theta\Delta t \gg 1$ in Eq.~\ref{eq:44},
 corresponding to homogeneous or life-time limited broadening. 
 We also have to conclude that the Jensen inequality can only be applied in 
 one direction since starting from the
 assumption that $\gamma_t$ is a mean-reverting OU process gives the 
 results presented in the previous section. 
This also would imply that care must be taken in interpreting the {\em stationary} lineshapes since the 
two different models for the background process 
appear to give similar spectral signatures.

If we specify the initial value of the coupling $\gamma(0)$ at $t=0$, we can use Jensen's inequality to compute an 
upper limit of the covariance as
\begin{align}
    \langle\gamma(t),\gamma(s)\rangle 
    \leq &\frac{\sigma\omega^2}{\sqrt{2\theta}}\left[e^{-\theta|t-s|}-e^{-\theta(t+s)}\right]^{\frac{1}{2}}.
\end{align}
If we take both $t$ and $s$ at
some later times such that the memory of the initial condition 
is lost and take $\Delta t = t-s$, we recover  Eq.~\ref{eq:stationary-cov} 
as the stationary covariance. 

In the non-stationary limit, however, the time-evolution of the mean (and hence $g_1(t)$) is very different.  Using Mathematica, we were able to arrive 
at an analytical expression for $\langle \gamma(t)\rangle$ as given in the Appendix.  Unlike its counterpart in the previous section, it does not 
relax exponentially to a stationary value and the 
resulting long-time value is far more complex. 
This suggests that one needs to look at both the 
line-shape and its temporal evolution to correctly
extract the background dynamics. 

\end{widetext}

\section{Discussion}

In this paper, we further explore how the spectroscopic
lineshape function reveals details of the 
electronic environment of an exciton. In particular,
we consider the effect of pair excitations that arise 
from the full many-body expansion of the 
Bosonic Hamiltonian.  As a simplifying assumption, 
we make the \textit{ansatz} that these can be 
treated as a single classical  variable 
that satisfies a known stochastic process, in this 
case the Ornstein-Uhlenbeck or Brownian motion process. 
We find that this does produce the lineshape function
given by the Anderson-Kubo model, albeit with twice the
coherence time.  Importantly, we show that the model
captures the formation
of exciton/polarons as the steady-state/long-time limit.  
We also consider the reversed case where we assume 
that the frequency fluctuation obeys the Anderson-Kubo model and derive the underlying SDE for the 
pair-fluctuations and their spectral density. 
In this latter case, we show that 
working backwards from the 
stochastic frequency dynamics one can recover the 
the underlying 
spectral density in the long-time limit using the 
Jensen inequality. 

The results from Sec.\ref{Sec:2.1} can be generalized for the 
case where the background process governing $\gamma(t)$ has a
known spectral density that can be expressed as a 
series of exponential functions. Under this case, the
$\gamma(t)$ is a sum over independent Ornstein-Uhlenbeck processes
and thus, $z(t)$ will be a sum over independent processes. 
On the other hand, 
as discussed in Sec.\ref{Sec:2.2},
inverting from an assumed process for $z(t)$
is non-trivial even for the rather simple model 
presented here; however, we 
can find an upper limit for the background spectral density governing $\gamma(t)$.  
\textcolor{erb}{While this 
can be taken simply as a mathematical exercise, we learn
is that one can obtain from the 
lineshape a bound on the spectral density given 
a model system/bath interaction, but not the exact spectral density in all but specific cases. 
This is potentially a useful result for developing machine learning methods for spectral analysis.
}

We also believe that the approach can be extended to 
account for quantum noise effects by treating the $q\ne 0$ 
terms in Eq.\ref{modelH} as quantum noise terms rather than treating them as a collective classical variable.  
Furthermore, while our current approach is limited to bosonic excitations, it is possible to extend this approach to 
fermionic systems at finite temperature.

\textcolor{marked}{
The cumulant expansion is a convenient technique to evaluate a function of random variables. For example, Bicout and Szabo's work \cite{BicoutJCP1998} provides a numerical strategy to compute the cumulant to arbitrary orders for a quadratically transformed stochastic process, namely passage through a fluctuating bottleneck, in its equilibrium/stationary distribution. 
It may be possible to extend our model to account for  non-Markovian dynamics using the projection of multidimensional Markovian processes
using their approach.
Here, 
we truncate this expansion at 
the second cumulant, which is 
sufficient  according to the Marcienkiewicz theorem~\cite{Marcinkiewicz1939,MarcinkiewiczTheorem:PRA1974}.  We emphasize that the non-stationarity of the 
dark-exciton background, and 
its dynamical coupling to the 
bright states, is a central 
component of the work presented here 
and could not be accounted for using
a stationary picture as in Ref.~\citenum{BicoutJCP1998}.
}

\appendix

\begin{acknowledgments}

The work at the University of Houston was funded in
part by the  National Science Foundation (CHE-2102506) and the Robert A. Welch Foundation (E-1337). 
The work at LANL was funded by Laboratory Directed Research and Development (LDRD) program, 20220047DR. The work at Georgia Tech was funded by the National Science Foundation (DMR-1904293). %

\end{acknowledgments}

{\bf Data Availability:}The data that supports the findings of this study are available within the article.


%

\begin{widetext}
\section{An alternative way to characterize $z(t)$}
\label{sec:gaussian}
\label{app:A}

In Sec.\ref{Sec:2.1} we defined a 
transformed statistical process $z(t)=\gamma(t)^2/\omega^2$ to describe the relative frequency shift, which is attributed to the bi-excitation interaction characterized by $\gamma(t)$. 
Therefore, in principle, the statistical property of $z(t)$ can be derived from  that of $\gamma(t)$. In this section, we will illustrate this approach by taking $\gamma(t)$ as an Ornstein-Uhlenbeck process, which is both Gaussian and Markovian. In the text we took the advantage of the Wiener process, using the It{\^o} calculus, to solve $z(t)$ and its statistical property from the SDE. Here, we will make use of the statistical property of $\gamma(t)$ to find the mean value and autocorrelation function of $z(t)$, which are higher moments of $\gamma(t)$, using the property of the multivariate Gaussian distribution, without solving the SDE. 

Let ${\bm X}=\{x_1,x_2,...,x_N\}^{\rm T}$ denote a set of Gaussian random variables with mean value vector ${\bar{\bm X}}=\{{\bar x_1},{\bar x_2},...,{\bar x_N}\}^{\rm T}$. The $k$-th moments read
\begin{equation}
    \left<\prod_{j=1}^N \left(x_j-{\bar x}_j\right)^{r_j}\right> =
        \begin{dcases*}
            0, &if $k$ is odd, \\
 \frac{(2\lambda)!}{\lambda! 2^{\lambda}} \left\{\sigma_{jk}\sigma_{mn}...\right\}_{\rm sym}, &if $k$ is even,
        \end{dcases*}
\end{equation}
where $k=\sum_j r_j=2\lambda$. $\{...\}_{\rm sym}$ denotes the symmetric bilinear form on the Gaussian vector space. Being specific, it is the sum of the product of $\sigma$'s allocated into $\lambda$ pairs. The set of \{$\sigma_{jk}=\langle x_j,x_k\rangle\}$ forms the covariance matrix.

The Ornstein-Uhlenbeck process $\gamma(t)$ has the expected value 
\begin{equation}
    \langle\gamma(t)\rangle = \gamma_0e^{-\theta t},
\end{equation}
where the initial average is $\gamma_0=\langle\gamma(0)\rangle$. We consider the Ornstein-Uhlenbeck process with indeterministic initial condition, therefore the covariance reads
\begin{align}
    \langle \gamma(t),\gamma(s) \rangle = \sigma_{\gamma_o}^2 e^{-\theta(t+s)} + \frac{\sigma^2}{2\theta}\left[e^{-\theta|t-s|}-e^{-\theta(t+s)}\right],
\end{align}
where $\sigma_{\gamma_o}^2$ describes the fluctuation at time zero.

Before proceeding to the property of $z(t)$, we take a careful consideration about its initial condition. Since the average $z_0=\langle z(0)\rangle=\langle \gamma(0)^2\rangle/\omega^2$, one has
\begin{equation}
    \omega^2 z_0 = \sigma_{\gamma_o}^2 + \gamma_0^2.
    \label{eqn:z0}
\end{equation}
Assuming the initial distribution of $\gamma(0)$ is Gaussian, one can use its fourth moment to find the fluctuation of $z(0)$
\begin{equation}
    \omega^4 \sigma_{z_o}^2 = 2\sigma_{\gamma_o}^4 + 4\sigma_{\gamma_o}^2 \gamma_0^2.
    \label{eqn:z0-fluc}
\end{equation}

The average of $z(t)$ can be found as
\begin{align}
    \left< z(t)\right> &= \frac{1}{\omega^2}\left< \gamma(t)^2\right> \nonumber \\
    &= \frac{\sigma_{\gamma_o}^2 + \gamma_0^2}{\omega^2} e^{-2\theta t} + \frac{\sigma^2}{2\theta \omega^2} \left(1 - e^{-2\theta t}\right),
\end{align}
which is exactly same as in  Eq.~(\ref{eqn:z-mean}) upon  substitution of Eq.~(\ref{eqn:z0}). Similarly, one finds the covariance
  \begin{align}
    \left<z_(t),z(s)\right> =& \frac{2\sigma_{\gamma_o}^4+4\sigma_{\gamma_o}^2\gamma_0^2}{\omega^4} e^{-2\theta(t+s)} + \frac{\sigma^4}{2\theta^2\omega^4} \left[e^{-\theta|t-s|} - e^{-\theta(t+s)}\right]^2 + \frac{2\sigma^2}{\theta\omega^4}\left(\sigma_{\gamma_o}^2 + \gamma_0^2\right) e^{-\theta(t+s)} \left[e^{-\theta|t-s|} - e^{-\theta(t+s)}\right].
  \end{align}
Substituting Eqs.~(\ref{eqn:z0}) and (\ref{eqn:z0-fluc}),
one finds the the covariance given here to be identical as in Eq.(\ref{eqn:zt-cov}).

\section{Changing variables using the
It{\^o} identity}
\label{app:B}

We briefly review the change of variable
procedure under It{\^o} calculus.  In 
general, we write a stochastic process as
\begin{align}
{\rm d}x = A[x]{\rm d}t + B[x]{\rm d}W_t    
\end{align}
where $A[x]$ and $B[x]$ are both independent functions of the 
stochastic variable $x$
and time $t$ and $W_t$ is the Wiener process with ${\rm d}W_t^2 = {\rm d}t$ 
according to the It\^o identity.
Often, we need to cast a function the stochastic variable, $f[x(t)]$,
in the form of an It{\^o} stochastic equation. For this
we need to perform a change of variables
\begin{align}
    {\rm d}f[x] &= f[x+dx]-f[x]
    \nonumber
    \\
    &=f' {\rm d}x + \frac{1}{2}f'' {\rm d}^2\gamma + \cdots \nonumber 
    \\
    &= f' \left(A[x]{\rm d}t+B[x]{\rm d}W\right) + \frac{1}{2}f''  \left(A[x]{\rm d}t+B[x]{\rm d}W\right)^2 \nonumber 
    \\
    &=(f' A[x] + \frac{1}{2}f''B[x]^2){\rm d}t
    + f' B[x]{\rm d}W_t
\end{align}
where $f'$ and $f''$ denote partial derivatives of $f[x]$
respect to $x$.  Under this, $f[x(t)]$ is considered as a transformed
process with respect to the original $x(t)$ stochastic variable.
It is straightforward to generalize this approach for vectors including
correlation between stochastic terms.  The reader is referred to 
Gardner's excellent book for more details on stochastic methods and their applications.\cite{Gardner}

\section{Expression for $\langle \gamma(t)\rangle$ from 
Sec.\ref{Sec:2.2}}
\label{app:C}

We give here the expression derived for the expectation value of
\begin{align}
    \gamma(t)=\pm \omega \sigma^{1/2} \left[\int_0^t e^{-\theta(t-\tau)} {\rm d}W_{\tau}\right]^{1/2}
\end{align}
which is the solution of the It{\^o} SDE
\begin{align}
    {\rm d}\gamma_t = - \left(\frac{\theta}{2}\gamma_t + \frac{\sigma^2 \omega^4}{8} \gamma_t^{-3}\right) {\rm d}t + \frac{\sigma\omega^2}{2}\gamma_t^{-1} {\rm d}W_t.
\end{align}
Recall, that this is a transformed process in which we 
assumed that the observed frequency fluctuations 
were from an Ornstein-Uhlenbeck process with $\gamma(t)=\omega\sqrt{z(t)}$.
Taking the initial value to be $z(0)=0$,
one finds the average as
\begin{align}
    \langle\gamma(t);z(0)=0\rangle =
 \gamma_{eq}
\frac{1}{(1-e^{-2 \theta  t})^{1/2}
(\coth (\theta  t)+1)^{3/4}}
\end{align}
with 
\begin{align}
\gamma_{eq}=
\frac{\left(\frac{1}{2}+\frac{i}{2}\right) \omega  }{\sqrt[4]{\theta }}
   \sqrt{\frac{\sigma}{\pi}}
   \Gamma \left(\frac{3}{4}\right)
\end{align}
which gives the red-shift of the exciton/polaron
energy due to the pair-fluctuations. 
The $z(0)=0$ initial condition is of course a special 
case. Using {\em Mathematica}, one can arrive at a general expression 
for $\langle\gamma(t)\rangle$;
however, as the expression is long and complicated
we will not reproduce it here.
The fact that $\gamma_{eq}$ can 
be complex-valued poses no
difficulties since, formally, 
we can equivalently write  the Hamiltonian
 in Eq.\ref{eq:2} as
\begin{align}
    H = \hbar\omega a^\dagger a + \gamma(t)a^\dagger a^\dagger + \gamma^*(t)
    aa.
\end{align}
with eigenvalues
\begin{align}
\lim_{t\to \infty}
\langle \tilde\omega\rangle&
\approx \sqrt{\omega^2 - |\gamma_{eq}/\hbar|^2}.
\end{align}
It is important to point out that
that while the eigenvalue $\tilde\omega(t)$ 
depends upon $\gamma(t)$, we have already specified
its evolution via the linearization expansion
in Eq.\ref{eq:w-eigen}
and by specifying $z(t)$ to be an Ornstein-Uhlenbeck 
process. What we find instead is that
$\langle\gamma(t)\rangle$ 
relates to the physical evolution of the 
background fluctuation that can be inferred 
by assuming a specific spectral model for the 
line-shape. 


\color{hao}
\section{Expression of correlation functions $R_{\alpha}$ in Section \ref{Sec:2.1}}
\label{app:D}

The correlation function Eq.(\ref{Eq:Rn}) can be written in terms of the stochastic variable $z(\tau)$ using Eq.(\ref{eq:w-eigen}). Carrying out cumulant expansion to the second order, we have
\begin{align}
    R_n(\tau_3,\tau_2,\tau_1) &= \left(\frac{i}{\hbar}\right)^3 \mu^4 \exp\left[i \omega \sum_{j=1}^3 (\pm)_j \tau_j\right]\ \exp\left[\sum_{n=1}^{\infty}\frac{(-i\omega/2)^n}{n!}\left<\sum_{j=1}^3(\pm)_j \int_0^{\tau_j} z(\tau) {\rm d}\tau \right>_{\rm c}\right] \nonumber \\
    &\approx \left(\frac{i}{\hbar}\right)^3 \mu^4 \exp\left[i \omega \sum_{j=1}^3 (\pm)_j \tau_j\right] \exp\left[-\frac{i\omega}{2}\sum_{j=1}^3 g_1(\tau_j)\right] \exp\left[-\frac{\omega^2}{8}\sum_{i,j=1}^3(\pm)_i (\pm)_j g_2(\tau_i,\tau_j)\right],
\end{align}
where $g_1(\tau)$ is given by Eq.(\ref{eq:g1-new}), and the two-time lineshape function is
 \begin{align}
    g_2(\tau_1,\tau_2) =& \int_0^{\tau_1} \int_0^{\tau_2} \langle z(\tau),z(\tau')\rangle {\rm d}\tau{\rm d}\tau' \nonumber \\ 
    =& \frac{\sigma_{z_o}^2}{4\theta^2}\left(1-e^{-2\theta \tau_1}\right)\left(1-e^{-2\theta \tau_2}\right) + \frac{\sigma^2 z_0}{2\theta^3\omega^2} \left\{1-e^{-2\theta(\tau_1+\tau_2)}-2e^{-2\theta \min(\tau_1,\tau_2)} + [1-2\theta\min(\tau_1,\tau_2)] \left(e^{-2\theta \tau_1}+e^{-2\theta \tau_2}\right)\right\} \nonumber \\
    &+ \frac{\sigma^4}{8\theta^4\omega^4}\left[e^{-2\theta(\tau_1+\tau_2)}-e^{-2\theta|\tau_2-\tau_1|}+4e^{-2\theta\min(\tau_1,\tau_2)}-4 +4\theta\min(\tau_1,\tau_2)\left(e^{-2\theta \tau_1}+e^{-2\theta \tau_2}+1\right)\right].
\end{align}

\textcolor{erb}{
\section{Concerning the unitary transformation 
to diagonalize Eq.\ref{eq:13}}
In our derivations we employed a very useful technique 
to bring the Hamiltonian in Eq.~\ref{eq:13} to diagonal form.  We give a 
quick review here for the interested reader. 
Writing $H$ in terms of diagonal and non-diagonal terms
\begin{align}
    H = H_o + W
\end{align}
one can easily show one can transform $H$ into a 
diagonal via
\begin{align}
    \tilde H = e^{-S}H e^{S}
\end{align}
in which the operator $S$ is related to $H_o$ and $W$ via
\begin{align}
    [H_o,S] = -W.\label{AD-condition}
\end{align}
This last step is obtained by expanding the exponential. 
Having $S$, one now defines the transformed operators
$\tilde a$ and $\tilde a^\dagger$ using a equations of motion
approach such that
\begin{align}
    a(\tau) = e^{-S \tau} a e^{S \tau}
\end{align}
with $a(0) = a$ and $a(1) = \tilde a$ for all operators. 
Taking derivatives, one obtains Heisenberg equations
\begin{align}
    \partial_\tau a(\tau) = [a, S]
\end{align}
which can be integrated to give the results and transformed 
variables in the paper.  For the case at hand, 
with $H_o = \hbar \omega (a^\dagger a + 1/2)$,
$W = \hbar\gamma(a^\dagger a^\dagger + a a)/2$, and $S = \xi(a^\dagger a^\dagger-aa)/2$ satisfies the condition in
Eq.~\ref{AD-condition} and one obtains
\begin{align}
    \partial_\tau a(\tau) &= -\gamma(t) a^\dagger(\tau) \\
    \partial_\tau a^\dagger(\tau) &= -\gamma(t) a(\tau) 
\end{align}
Note that $\gamma(t)$ is time-dependent, the variable $\tau$ is not chronological time, it is simply introduced
at each chronological-time and does not interfere 
with the time-ordering or  integration over the 
stochastic variable $\gamma(t)$.  Requiring the transformed
$\tilde H$ to be diagonal yields the results around Eq.~\ref{eq:13}.
}

\end{widetext}

\end{document}